\documentclass[fleqn,twoside]{article}
\usepackage{espcrc2}

\usepackage{graphicx}
\usepackage[figuresright]{rotating}

\newcommand{\AmS}{{\protect\the\textfont2
  A\kern-.1667em\lower.5ex\hbox{M}\kern-.125emS}}

\title{Bubbling solutions, entropy enhancement and the fuzzball proposal}

\author{C.Ruef\address[IPhT]{Institut de Physique Th\'eorique, \\ 
        CEA Saclay Orme des Merisiers 91191 Gif-sur-Yvette}%
	}

\begin{document}

\begin{abstract}
In this short note we explain the main idea of the work done in \cite{Bena:2008nh,Bena:2008nh2}. We present a family of black hole microstates, the bubbling solutions. We then explain how supertubes placed in such backgrounds have their entropy enhanced by the presence of the background dipole charges. This indicates this could account for a large amount in the entropy of the three charge black hole.
\vspace{1pc}
\end{abstract}

\maketitle

\section{Introduction}

It is a well-known fact that every black hole has an entropy, proportional to the area of its horizon. But the microscopic origin of this entropy is still unclear, despite the fact that much progress has been done in this direction over the last decades, in particular in the context of string theory. Indeed, Strominger and Vafa were able to compute the entropy of a class of supersymmetric (SUSY) black holes from a microscopic point of view \cite{Strominger:1996sh}, by counting bound states of strings and branes at small effective coupling constant. SUSY then protects the counting when this coupling grows, and the entropy matches the macroscopic one at large coupling, when the black hole exists. So this counting, despite its success, doesn't tell us anything about the nature of a ``black hole microstate" at large coupling.

Another very useful tool to count the microscopic entropy of black holes is the $AdS$/CFT correspondance. If one wants to describe a black hole in $AdS$ space, or a hole in flat space but with an $AdS$ throat in its near-horizon limit, the correspondance allows one to work in the dual CFT, where it is easier to count the number of states. Once again this approach, despite its success, doesn't explain what is a black hole microstate. 

One way to answer this question is the so called fuzzball proposal, proposed by Mathur (see for ex. \cite{Mathur:2005zp}). It states that every black hole microstate is related to a horizonless solution, which can be very complicated, very fuzzy, but stay nevertheless completely regular and look like a black hole outside the (would be) horizon. We also don't know{\it{a priori}} if the supergravity (SUGRA) regime is enough to describe all this microstates or if they are fully stringy. The black hole should be seen in this picture as a coarse grained description of these geometries. This proposal has been successfully verified in the case of the two-charge black hole. However, the two-charge black hole does not have classically a real macroscopic horizon, one has to define a ``stretched horizon", with entropy $2\pi\sqrt{2Q_1Q_2}$, the $Q_i$'s being the charges of the back hole. Different groups have been able to construct and count the corresponding microstates, in the context of SUGRA, and to reproduce the correct entropy $2\pi\sqrt{2Q_1Q_2}$ \cite{Lunin:2002qf,Lunin:2002iz}. What remains to be done is to understand the three-charge case, where the 5D black hole has a real macroscopic horizon, with entropy $2\pi\sqrt{Q_1Q_2Q_3}$.

\section{Bubbling solutions}

One very interesting class of three-charge solutions is the one of the multicenter solutions, or bubbling solutions \cite{Denef:2002ru,Bena:2005va,Berglund:2005vb}. This class describes a very large spectrum of different black hole configurations: black holes, black rings, concentric black rings or black saturns as well as smooth horizonless geometries. Let's concentrate on the smooth solutions. In this type of configurations, the main idea is that the external space isn't flat anymore but topologically non trivial: there are a lot of Taub-NUT like centers, which all carry electric and magnetic charges. This creates non contractible two-spheres, or bubbles, held by the fluxes between the centers. The electric charges seen from infinity are then created by these magnetic fluxes. The smoothness of the solution is ensured by the bubble equations - one for each center - relating the charges of the centers and the distance between them 

\begin{equation} \label{bubble}
 \sum_{j,j\neq i}\frac{<\Gamma_i,\Gamma_j>}{r_{ij}}=<h,\Gamma_i>,
\end{equation}
where $i$, $j$ denotes the different centers, $r_{ij}$ is the distance between them, $\Gamma_i$ encodes the charges of the center $i$ and $h$ the charges at infinity, and $<\,,\,>$ is a symplectic product between the centers, representing the interactions. Since it will be useful in the next section, let us also mention that this bubbling solutions do not have just usual electric charges but also magnetic dipole moments, invisible from infinity but which will play an important role in the following. 


\section{Entropy enhancement}

Supertubes \cite{Mateos:2001qs} are very interesting objects, particularly in this context of constructing smooth SUGRA solutions. They are supersymmetric brane configurations with two electric charges and one magnetic dipole charge, and can have classically arbitrary shapes. Ignoring the backreaction on the environnement, one can use them to probe a certain background. This arbitrary shape gives us an infinite dimensional moduli space, which  becomes finite dimensional after quantization and gives us an entropy $S\sim\sqrt{Q_1Q_2}$, when the tube is in flat space. Here is now the interesting issue: if one puts such a tube in a background with dipole charges, like a bubbling background, the entropy is enhanced to $\sqrt{Q_{1\mathrm{eff}}Q_{2\mathrm{eff}}}$ \cite{Bena:2008nh}. The effective charges appearing in the entropy contain two contributions, one from the usual electric charge and one from the dipole-dipole interaction. The entropy then depends on the electric charges and magnetic dipole moments of both the tube and the background and also on the shape of the tube and its position in the background. Playing with this elements, we can have this entropy becoming very large and eventually behaving like $\sqrt{Q^3}$, where $Q$ is a typical value of the electric charges. In other words, we expect that this could account for a large fraction of the black hole entropy.

\medskip

Others works \cite{Bena:2008nh2,Bena:2008wt} shed a new light on this result. In \cite{Bena:2008wt} we learn that through spectral flow, each center of a bubbling configuration can be turned into a supertube. So the previous analysis is in fact more general: each center can be dualized into a supertube and can have its entropy enhanced. Finally, in \cite{Bena:2008nh2}, we compare the probe Born-Infeld approach and the full backreacted SUGRA approach. We learn that the probe approach captures all the physics of the problem and confirm that the probe computation of \cite{Bena:2008nh} will give rise to smooth SUGRA solutions.

\end{document}